\documentclass[epj,nopacs]{svjour}
\usepackage{graphics}
\usepackage{turnstile}
\usepackage{graphicx,float}
\usepackage{amsmath,amssymb}
\usepackage{url}
\usepackage[labelformat=simple]{subcaption}
\usepackage{epsfig}
\usepackage{epstopdf}
\usepackage{widetext}
\usepackage{cite}

\begin{document}
\title{
			Interaction of a wave packet with potential structures moving with acceleration
		}
\author{M.~A.~Zakharov\inst{1,2} \thanks{\email{ zakharovmax@jinr.ru}} \and G.~V.~Kulin\inst{2} \and A.~I.~Frank\inst{2}}                     % Do not remove
\institute{Moscow Institute of Physics and Technology, 141701 Dolgoprudny, Russia \and Joint Institute for Nuclear Research, 141980 Dubna, Russia}
\date{\today}
\abstract{
The paper is devoted to a numerical study of the problem of interaction of the wave packet with potential structures moving with constant acceleration. In all the cases considered the result of the interaction is a change in the velocity spectrum. In the first approximation the magnitude of the shift in the spectrum is determined by the product of acceleration by group delay time. Also, as the direction of acceleration reverses the effect changes its sign. The results are completely consistent with the idea of universality of the Effect of Acceleration which consists in a change in the frequency of the wave at scattering on an object moving with acceleration.
%
%\PACS{
   %  {PACS-key}{discribing text of that key}   \and
   % {PACS-key}{discribing text of that key}
  % } % end of PACS codes
} %end of abstract

\maketitle
\section{Introduction}
	\label{sec:intro}
	
	The study of the optical phenomenon later called the accelerating matter effect has a rather long history. It was apparently first mentioned in the PhD thesis of Mikerov~\cite{Mikerov1977} who found by numerical calculation that the energy of ultracold neurons (UCNs) must change after their passage through a foil oscillating in space. Later, Tanaka found a solution to the problem of passage of an electromagnetic wave through a dielectric plate moving with constant acceleration. He showed that the frequency of the wave that passed through such a sample differed from that of an incident wave~\cite{Tanaka1982}. With a single passage of the wave through the plate the frequency change is determined as follows
	
	\begin{equation}\label{eq1}
	\Delta \omega = \frac{\omega a d}{c^{2}} (n - 1), \;\;\; \left( \frac{a d}{c^{2}} << 1\right).
	\end{equation}
	
	Here $\omega$ is the incident wave frequency, $n$ is the refractive index, $d$ is the sample thickness, $a$ is the acceleration, and $c$ is the speed of light. The possibility of observing such optic phenomenon is discussed in the work~\cite{Neutze1998}. 
	
	%Ten years
	A decade after publication of~\cite{Tanaka1982} Kowalski discussed the possibility of testing the equivalence principle in a new type of neutron experiment and came to the conclusion that the energy of the neutron would change following its passage through an accelerating refracting sample~\cite{Kowalski1993}. The energy change was determined as
	
	\begin{equation}\label{eq2}
	\Delta E = m a d \left(\frac{1-n}{n}\right),
	\end{equation}
	where $m$ is the neutron mass. 
	
	This problem was also considered in a later work by Nosov and Frank~\cite{Nosov1998}. The analysis consisting in carrying out sequential calculation of the velocity of the neutron on entering the sample, on propagating in some accelerating medium, and on escaping it through its other surface was in fact based on a classical approach. For the effect, the authors obtained the equation that completely coincided with that previously obtained by Kowalski~\eqref{eq2}.
	
	Experimental observation of the effect of UCN energy change on passage through a sample oscillating in space was reported in~\cite{Frank2006,  Frank2008}. For the energy transfer of a fraction of neV there was obtained good agreement with Eq.~\eqref{eq2}. In a later work~\cite{Frank2011} a change in the neutron velocity on passage through an oscillating plate of refracting substance was registered. The obtained results were also in satisfactory agreement with the expected ones.
	
	Relatively recently, a change in the energy of cold neutrons passing through an accelerating crystal under conditions close to those of Bragg reflection, was observed~\cite{Voronin2014,  Braginetz2017}. It is natural to assume that the effect detected in the works is of the same nature as the change in frequency at refraction.
	
	As regards the physical interpretation of the effect, different works demonstrate different theoretical approaches. While the corpuscular approach is used in~\cite{Mikerov1977,  Nosov1998}, the Kowalski method~\cite{Kowalski1993} bases on calculating the wave propagation time through a refracting sample in the laboratory system of reference and in that moving with acceleration.
	
	The optical approach dating back to~\cite{Kowalski1993} appeals naturally to the Doppler shift of the wave frequency at refraction into a moving medium. This approach finds development in the later works~\cite{Frank2013, Frank2016}. If refraction occurs at the boundary of a moving medium, the Doppler shift of the neutron wave frequency is                                    
	
	\begin{equation}\label{eq3}
	\Delta \omega_{d} = (n' - 1) k V, \;\;\; (V \ll v),
	\end{equation}
	where $n'$ is the refractive index in the coordinate system of moving matter, $v$ and $k$ are the velocity and the wave number of the incident wave, respectively, $V$ is the velocity of matter and of its boundary.
	
	In the case of a uniformly moving layer of matter the Doppler frequency shifts due to passage of the wave through the layer's two boundaries are equal in value but opposite in sign. The total effect is then equal to zero. 
	
	In the case of accelerated motion, the frequency shifts on the entrance and escape surfaces are different since the velocity of the boundary changes by $\Delta V = a t$   during the wave propagation through the sample. Here $t$ is the transit time of the sample. In the considered case of a refractive sample, the difference between the frequencies of the incident and transmitted neutron waves is                                                
	
	\begin{equation}\label{eq4}
	\Delta \omega_{d} = a d \frac{1-n}{n} \frac{k}{v} = \frac{mad}{\hbar} \frac{1-n}{n},\;\;\; (V, a t \ll v)
	\end{equation}
	and the energy change $\Delta E = \hbar \Delta \omega$ is determined by Eq.~\eqref{eq2}.
	
	However, it turns out that one can arrive at the conclusion that the frequency of the wave changes as a result of its propagation through a refracting sample without any particular optical calculation but just on the basis of the validity of the equivalence principle.
	
	It is demonstrated in~\cite{Frank2008} that the main factor that determines the magnitude of the effect is the difference $\Delta t$ in the times of wave propagation through free space and through the region of space where a refracting sample is located. Knowing the difference and neglecting both relativistic and non-stationary quantum effects one can unify the expressions for the effect for massive particles and for light in coincidence with those derived by Tanaka~\eqref{eq1} and Kowalski-Nosov-Frank~\eqref{eq2}.
	
	In both cases, the formula 
	
	\begin{equation}\label{eq5}
	\Delta \omega = k a \Delta t,
	\end{equation}
is correct, which allows us to come to the conclusion that we are speaking about some general optical effect for whose occurrence accelerating motion of a refractive sample is sufficient~\cite{Frank2011, Frank2016}. Moreover, the wave can be of any nature.

	In the recent work~\cite{Frank2020}, attention was drawn to an important fact that the conclusion about the change of the frequency of the wave passing through an accelerating sample was made in~\cite{Frank2008} just on the assumption of different propagation times of the wave in the sample and in vacuum, but, in no way, on the assumption that the difference $\Delta t$ is due to refraction in matter. Consequently, this conclusion holds if, instead of a refracting sample, we speak about an object moving with acceleration and transmitting a signal with a delay in time. 
	
	It is known, for example, that in quantum mechanics interaction is inevitably associated with the time delay described in the first approximation by the so-called group delay time (GDT)~\cite{Bohm1951, Wigner1955}.
	
	\begin{equation}\label{eq6}
	\tau = \hbar \frac{d \varphi}{d E},
	\end{equation} 
here $\varphi$ is the phase of the amplitude of the wave function of the particle that experienced interaction, e.g., scattering, and $E$ is the energy.  
	
	If the prediction made in Ref.~\cite{Frank2020} is true, the result of interaction of a particle with any object moving with acceleration is a change in its energy and frequency. Here we are speaking about some effect that complements the conventional Doppler shift and is proportional not to speed, but to acceleration. It can be expected that in this case, the frequency change is qualitatively described by Eq.~\eqref{eq5} where the role of time delay is played by GDT~\eqref{eq6}. 
	
	This paper is devoted to verification of the above mentioned hypothesis. Leaving aside the scattering and refraction of light, we restrict ourselves to the case of a massive nonrelativistic particle, having the neutron in mind. It seems that from the point of view of the efficiency of related experiments most promising is the use of slowest neutrons, i.e., UCNs. 
	
	It is known that interaction of slow neutrons with matter can be described by using the effective potential: 
	
	\begin{equation}\label{eq7}
	U = \frac{2 \pi \hbar^{2}}{m} \rho b,
	\end{equation} 
	where $m$ is the neutron mass, $\rho$ is the number of nuclei per unit volume, $b$ is the average bound coherent scattering length. In principal, it is possible to construct various configurations of one-dimensional potential structures by combining the layers of matter with different scattering length densities,  $\rho b$.
	
	We present the results of a numerical study of the interaction of UCNs with a number of simple quantum objects moving with acceleration. At the same time, we completely leave aside the question of the feasibility of the corresponding experiments.
	
	\section{Calculation technique}
	\label{sec:Calculation technique}
	
	The problem of interaction between the wave and a moving potential structure corresponds to the time-dependent Schrodinger equation
	
	\begin{equation}\label{eq8}
	i\hbar \frac{\partial}{\partial t} \Psi (x, t) = \left(-\frac{\hbar^{2}}{2m} \frac{d^2}{dx^2}+ V(x, t)\right)\Psi(x, t),
	\end{equation} 
	where $V(x, t)$ is some potential structure moving as a whole. To solve equation~\eqref{eq8}, we applied the same method as in paper~\cite{Zakharov2020}.  
	
	For some instant of time, $t'$, the solution can be presented as the result of action of the evolution operator $\hat{T}(x,t',t)$ on the wave function at the preceding instant of time, $t$~\cite{Hardin1973}. 
	
	The action of the operator $\hat{T}(x,t',t)$  can be found by numerical calculation based on the method of evolution operator splitting~\cite{Messiah1961}. The time interval during which it is necessary to trace the wave packet evolution is divided into short intervals, $\theta$. The solution for any time instant $t+\theta$ can then be presented as the result of action of the evolution operator $\hat{T}(x,t',t)$ on the wave function determined at the previous instant of time $t$:
	
	\begin{equation}\label{eq9}
	\Psi (x, t + \theta) = \hat{T}(x, t + \theta, t)\Psi(x, t).
	\end{equation} 
	
	According to works~\cite{Magnus1954, Fisher1973, Agrawal2001, Weiss1962}, the operator $\hat{T}(x, t + \theta, t)$ can be written in the form: 
	
	\begin{equation}\label{eq10}
	\hat{T}(x, t + \theta, t) = \hat{P}(x, t + \theta)\hat{K}\hat{P}(x, t),
	\end{equation}  
	where 
	
	\begin{equation}\label{eq11}
	\begin{split}
	\hat{P}(x, t) = \exp\left(-\frac{V(x, t)}{\hbar}\frac{\theta}{2}\right),\\ \hat{K} = \exp\left(i\theta\frac{\hbar}{2m}\frac{\partial^{2}}{\partial x^{2}}\right).
	\end{split}
	\end{equation}  
	
	The evolution of the wave function over the time interval $\theta$ can be then written as: 
	
	\begin{widetext}
	\begin{equation}\label{eq12}
	\Psi(x, t + \theta) = \exp \left(-i\frac{\theta}{2}\frac{V(x, t + \theta)}{\hbar}\right) \left\{\exp \left(i\theta\frac{\hbar}{2m}\frac{\partial^{2}}{\partial x^{2}}\right) \left[\exp\left(-i\frac{\theta}{2}\frac{V(x, t)}{\hbar}\right) \Psi(x, t)\right]\right\}+O(\theta^3).
	\end{equation}  
	\end{widetext}
	
	The action of the differential operator $\hat{K}$  on the function $A(x,t)$ can be easily calculated as follows
	
	\begin{equation}\label{eq13}
	\hat{K}A(x, t) = F^{-1}\{\hat{K}^{F}[FA(x, t)]\},
	\end{equation}  
where $F$ is the Fourier transform, $F^{-1}$is the inverse Fourier transform, $\hat{K}^{F}$  is the Fourier transform of the operator $\hat{K}$ : 
	
	\begin{equation}\label{eq14}
	\hat{K}^{F} = F\hat{K} =\exp \left(i\frac{\hbar}{2m}k^{2}\theta\right).
	\end{equation}  
	
	Therefore, numerical solution of the above-described problem is reduced to calculation of the wave function by the formula:
	
	\begin{widetext}
	\begin{equation}\label{eq15}
	\Psi(x, t + \theta) = \exp \left[ -i\frac{\theta}{2}\frac{V(x, t + \theta)}{\hbar} \right]
	F^{-1} \exp \left( i\theta\frac{\hbar}{2m}k^{2} \right) F \exp \left[ -i\frac{\theta}{2}\frac{V(x, t)}{\hbar} \right] \Psi(x, t).
	\end{equation}  
	\end{widetext}
	
	The calculation was performed sequentially for each time step. The fast Fourier transform algorithm (FFT)~\cite{Cooley1965} was used.
	
	\section{Interaction of a neutron with a potential structure moving with acceleration. Calculation results}
	\label{sec.Interaction of a neutron with a potential structure moving with acceleration. Calculation results}
	
	\subsection{Potential barrier}
	\label{subsec:Potential barrier}
	
	%The simplest example of the acceleration effect is probably 
	Consider the problem of interaction of the neutron with the potential barrier moving with constant acceleration. Continuous motion along the axis x significantly distinguishes the problem from the case of a potential oscillating in space considered in works~\cite{Pimpale1991, Zakharov2020}. In the present case, the potential is                               
	
	\begin{equation}\label{eq16}
	V(x,t) = U(x + a t^{2}/2),
	\end{equation} 
	
	\begin{equation}\label{eq17}
	U(x,0) =  \begin{cases}
	U_{0} &\text{0 $<$ x $<$ d}\\
	0 &\text{otherwise}
	\end{cases}
	\end{equation} 
	
	We restrict ourselves to the above-the-barrier case where the neutron energy exceeds the height of the barrier. The initial neutron wave function is in the form of the Gaussian wave packet: 
	
	\begin{equation}\label{eq18}
	\Psi_{0}(x) = \frac{1}{\sqrt[4]{\delta x^{2} \pi}}\exp \left(-ik_{0}x\right) \exp\left[-\frac{(x - x_{0})^{2}}{2\delta x^{2}}\right],
	\end{equation} 
where $\delta x$ stands for the width of the wave packet in coordinate space, $k_{0}$ is the neutron wave number that corresponds to the wave packet maximum. 
	
	\begin{figure*}[htb]
	\begin{minipage}[t]{.45\textwidth}
		%\centering

		\includegraphics[width=\columnwidth]{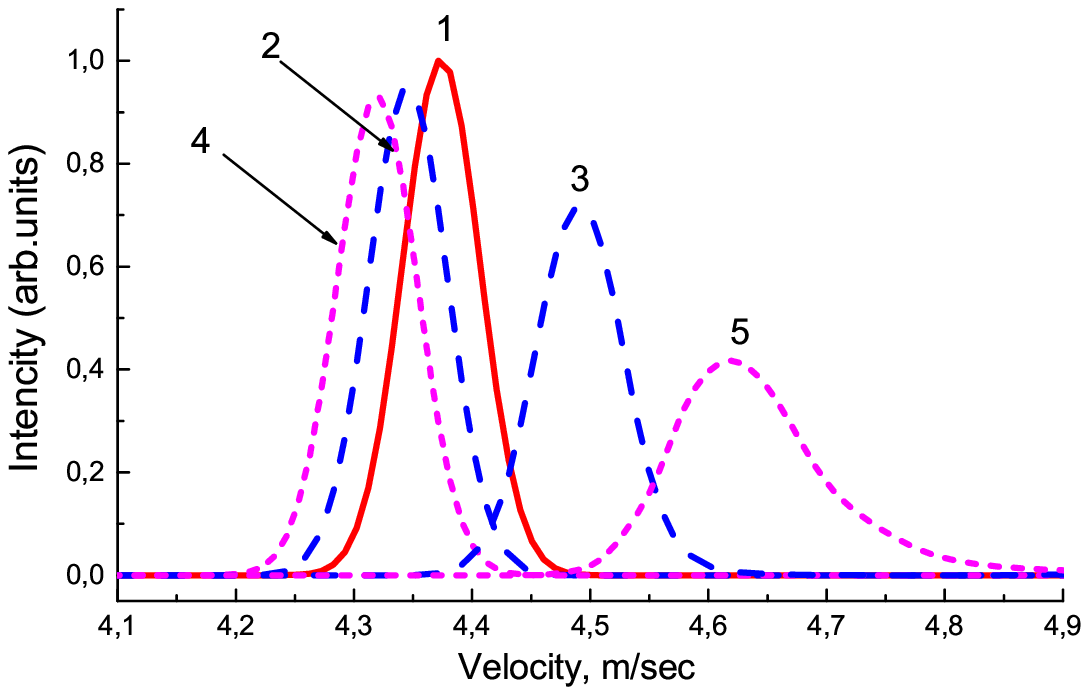}
		\caption
		{The velocity spectra of the UCNs that passed above the accelerating potential barrier vs the width of the barrier and the sign of acceleration. 1 - initial spectrum, 2 - 1 $\mu \text{m}$ (-), 3 - 1 $\mu \text{m}$ (+), 4 - 2 $\mu \text{m}$ (-), 5 - 2 $\mu \text{m}$ (+).}
		\label{Fig1}
	\end{minipage}
	%\end{figure}
	\quad
	%\begin{figure}[htb]
	\begin{minipage}[t]{.42\textwidth}
		%\centering
		\includegraphics[width=\columnwidth]{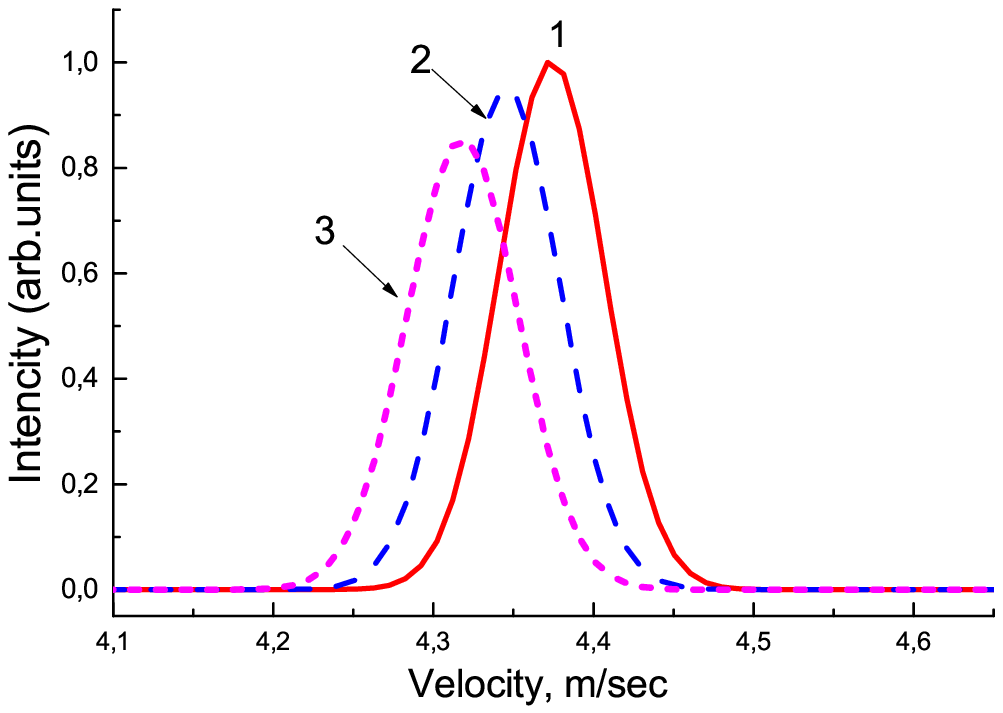}
		\caption
		{The velocity spectra of the UCNs that passed above the accelerating potential barrier vs the height of the barrier. 1 - initial spectrum, 2 - U = 50 neV, 3 - U = 75 neV.}
		\label{Fig2}
	\end{minipage}
\end{figure*}
	
	In the calculation, the neutron energy was accepted to be $E = 100~\text{neV}$, which corresponded to the velocity $v \approx 4.4~\text{m/s}$. The spatial width of the wave packet $\delta x$  could not be taken too small to avoid too strong increase in its energy width. On the other hand, an increase in the dispersion in the coordinate space led to an increase of the coordinate interval over which the calculation was performed, which increased the calculation time. For the energy width of the packet a value of $\delta E = 2~\text{neV}$ was taken, which corresponds to its spatial width of $\delta x = 1.35~\mu \text{m}$. Note that such values were close to the parameters of the UCN beam used in real experiments~\cite{Frank2008, Kulin2016}.
	
	The initial position of the wave packet was $x = - 5~\mu \text{m}$. So, at the moment $t = 0$ it was almost completely out of the potential region. Calculation of the packet evolution after interaction continued until the distance between its peak and the boundary of the potential reached the value of  $3\delta x$. Under such conditions, the packet was almost completely beyond the region of the potential.
	
	Figure~\ref{Fig1} shows the calculated velocity spectra of the states resulting from above-the-barrier transmission through an accelerating potential barrier for two widths of the barrier, $d$, and two directions of acceleration. The barrier height is $U = 50~\text{neV}$. The width and the sign of acceleration are indicated in the figure caption. The sign $+$ means that acceleration is in the direction of neutron velocity. For clarity of the result and to facilitate the calculation conditions, the acceleration is taken sufficiently large, namely $|a| = 10^{6}~\text{m}/\text{s}^{2}$.
	
	Figure~\ref{Fig2} shows the dependence of the spectrum shift on the potential barrier height. The acceleration is in the direction opposite to that of neutron velocity.
	
	As it can be seen in Figure~\ref{Fig1} the neutron velocity decreases if the acceleration of the barrier is in the direction of neutron velocity and increases for that in the opposite direction. For equal absolute values of acceleration the speed increment is larger in the case when acceleration directed along the neutron velocity, that is in the case of uniformly acceleration motion. This is obviously due to the fact that in this case the time it takes the particle to traverse the region of the accelerating potential is longer then in the case of retarded motion. For the same reason, the effect increases with increasing height and width of the barrier.
	
\begin{figure}[htb]
		\centering
		\includegraphics[width=\columnwidth]{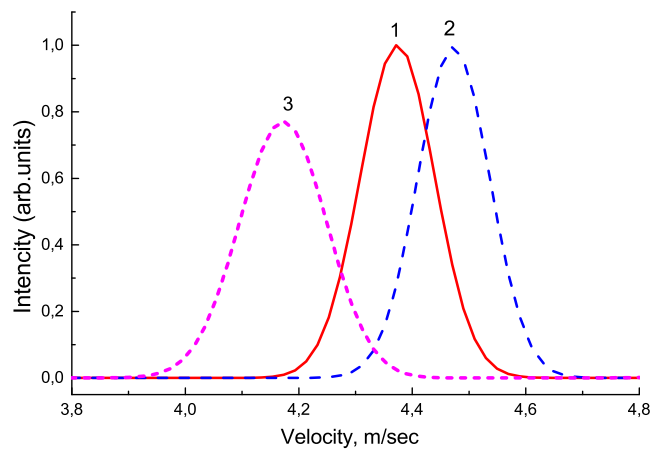}
		\caption
		{The velocity spectra of the UCNs that passes above the accelerating potential well. 1 - initial spectrum. 2 - acceleration of the potential in the direction opposite to neutron velocity. 3 - acceleration of the potential in the direction of the neutron velocity.}
		\label{Fig3}
	\end{figure}

	Figure~\ref{Fig3} shows the results of the calculation for the neutron that passes above the accelerating potential well. The depth of the well is 50 neV, its width is 1 $\mu \text{m}$. The absolute value of acceleration is $|a| = 10^{6}~\text{m}/\text{s}^{2}$.
		
	It can be seen that the sign of the effect is opposite to that for the barrier case because the group delay time becomes negative in this case~\cite{Muga2002}.
	
	Note that the above results are quite consistent with the concept of the accelerating matter effect~\cite{Frank2008} the equations for which were obtained from classical considerations. In fact, the region of the potential can be considered as some matter characterized by the refractive index                                               
	
	\begin{equation}\label{eq19}
	n = \sqrt{1-U/E}.
	\end{equation} 
	
	Since the particle energy used in the calculation was much higher than the potential and the barrier width was much larger than the neutron wavelength, the role played by quantum phenomena was relatively little.
	
\begin{figure}[htb]
		\centering
		\includegraphics[width=0.95\columnwidth]{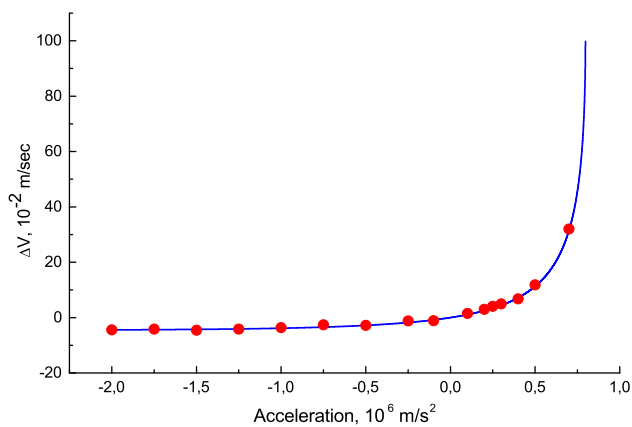}
		\caption
		{
			The change in the velocity of the neutron transmitted through the region of a potential moving with acceleration. The solid curve is a semi-classical calculation, the points are the quantum calculation of the wave packet evolution.
		}
		\label{Fig4}
	\end{figure}

	This fact is illustrated in Figure~\ref{Fig4}. It shows the results of calculation of the change in the velocity $d v$ during the transmission of the neutron through a potential barrier moving with acceleration in a wide range of acceleration. The results were obtained by the method described above~\eqref{eq15} and also, in the semi-classical approximation~\cite{Nosov1998} based on sequential calculation of the change in the neutron velocity at each boundary of matter characterized by some effective potential and, accordingly, by a refractive index.
	
	In the calculation the neutron energy was $E = 100~\text{neV}$, the value of the effective potential was $U = 50~\text{neV}$, and the width of the barrier was $d = 1~\mu \text{m}$. The results show good agreement of the numerical quantum calculation with those based on the semi-classical approach that is valid in this case.
	
	An increase in the effect at large positive accelerations is due to an increase in the neutron traversal time in the region of the potential. Since the right-hand boundary continues to move at an ever increasing speed during that time, the neutron needs more and more time to reach it due to growing acceleration. Starting with certain acceleration the neutron will not be able to reach the right-hand boundary at all as in some time it will find itself in free space having escaped through the left-hand boundary that has caught up with it~\cite{Nosov1998}.
	
\begin{figure}[htb] 
		%	\centering
		%	\captionsetup[subfigure]{labelformat=simple}
		%\hspace{-3\baselineskip}
		\captionsetup[subfigure]{labelformat=empty}
		\begin{subfigure}[b]{0.95\columnwidth}
			\caption{}
			\vspace{-1.5\baselineskip}
			\includegraphics[width=\textwidth]{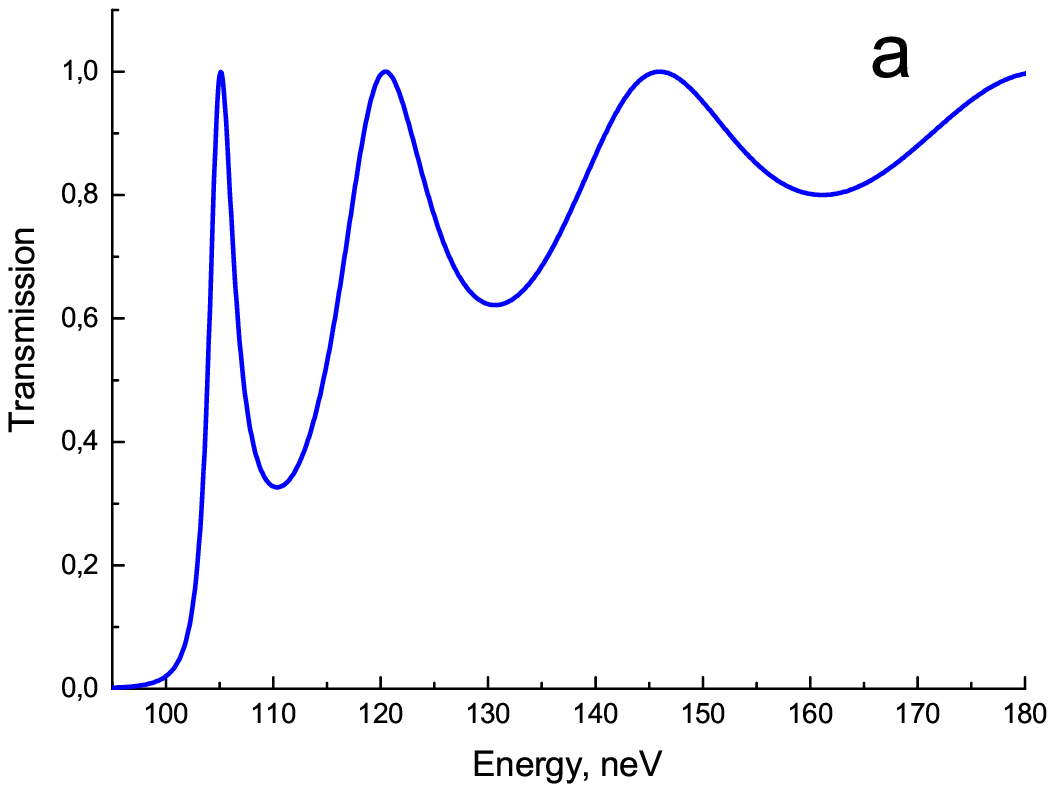}
			\label{Fig5a}
		\end{subfigure}
		%\vfill
		%	% \hspace{-0.6\baselineskip}
		\begin{subfigure}[b]{0.95\columnwidth}
			\caption{}
			\vspace{-2.5\baselineskip}
			\includegraphics[width=\textwidth]{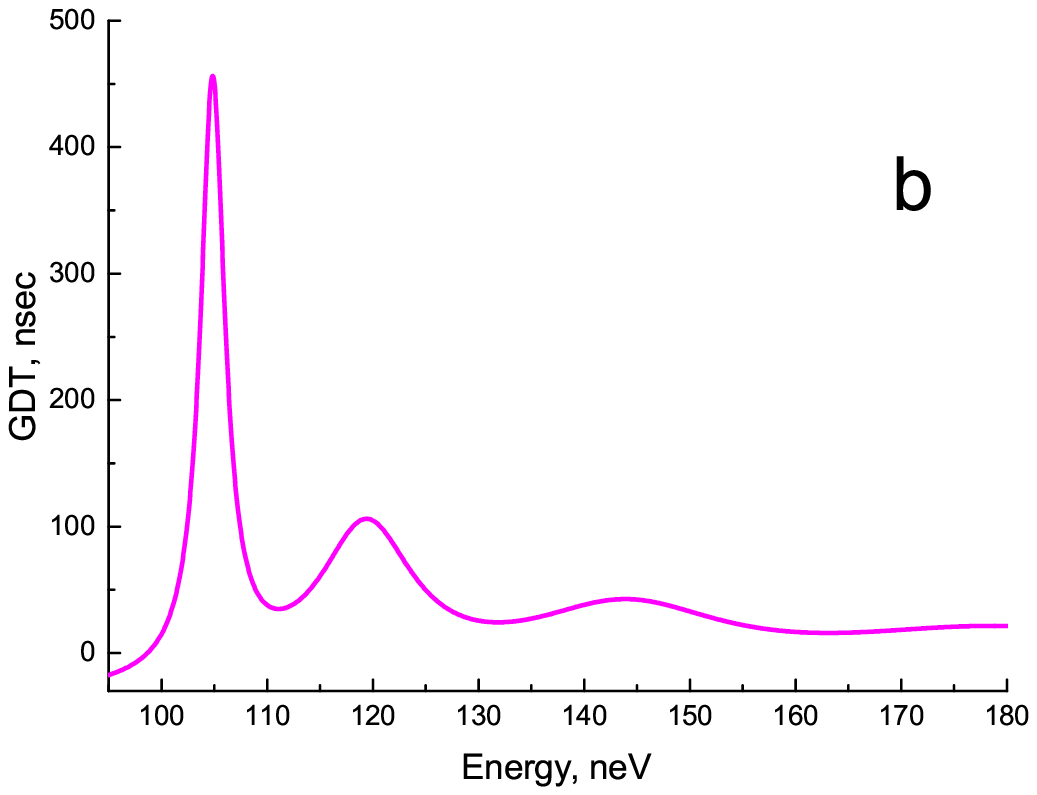}
			\label{Fig5b}
		\end{subfigure}
		%\vfill
		
		\begin{subfigure}[b]{0.95\columnwidth}
			\caption{}
			\vspace{-2.5\baselineskip}
			\includegraphics[width=\textwidth]{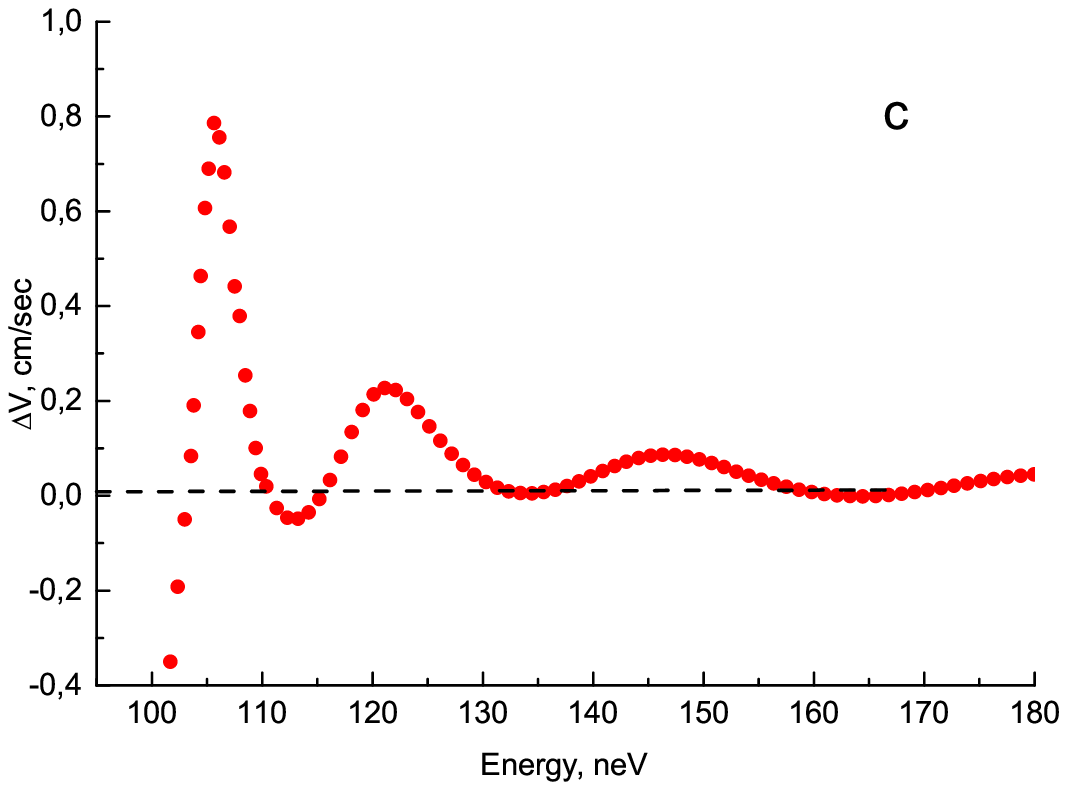}
			\label{Fig5c}
		\end{subfigure}
		%	\vspace{-0.7\baselineskip}
		\caption
		{The barrier transmission coefficients (a), the group delay time (b), and the change in neutron velocity $dv$ (c) vs neutron energy.}
		\label{Fig5}
	\end{figure}

	Since our main aim is to study the interaction of a particle with an accelerating potential in essentially quantum phenomena, we should turn to problems in which the role of quantum phenomena is enhanced. In particular, of interest is the case of a particle passing  above a barrier with a relatively small excess of its energy with respect to that of the potential $U$. It is obvious that as the energy decreases, the probability of reflection from the barrier borders increases as well as the role of interference effects, see Fig.~\ref{Fig5}.
	
	%\begin{figure}[htb]
	
%	Figure~\ref{Fig5} shows the barrier transmission, the group delay time, and the change in the neutron velocity due to the barrier accelerated movement in dependence on the neutron energy.
	
	The barrier height was $100~\text{neV}$, its width was $0.2~\mu \text{m}$ and the acceleration in the opposite to neutron velocity direction was equal to $2\times10^{4}~\text{m}/\text{s}^{2}$. It is obvious that in such conditions neutron deceleration is expected, which is confirmed by quantum calculations. A comparison of the curves in figure~\ref{Fig5} shows that the change of the neutron velocity qualitatively repeats the GDT curve behaviour, which is quite consistent with the prediction in Ref.~\cite{Frank2020}.
	
	Somewhat unexpected seems to be the negative value of the effect with an extremely small excess of energy  above the threshold as well as in the first minimum of transmission. Here, the points on the curve in figure~\ref{Fig5c} show acceleration of the neutron instead of the expected deceleration. This somewhat paradoxical result is related to a change in the shape of the wave packet passing through the region of the potential due to strong spectral dependence of the barrier transmission coefficient. The fact is that a significant part of the incident state is reflected under such conditions enriching the transmitted state with relatively fast components of the wave packet. Thus, the apparent neutron acceleration is explained by reduction of the wave packet having the same nature as the well-known Hartmann paradox~\cite{Muga2002, Hartman1962, Chen2008} or negative GDT at reflection of neutrons from an asymmetric potential structure~\cite{Bushuev2018}.
	
	\subsection{Neutron reflection from a potential step}
	\label{subsec: Neutron reflection from a potential step}
	
	Among the problems under consideration of particular interest is probably the interaction of a particle with an accelerating potential step. The group delay time at total reflection from a semi-infinite potential is~\cite{Frank2014}.
	
	\begin{equation}\label{eq20}
	\tau = \frac{\hbar}{\sqrt{E_{n}(U-E_{n})}},
	\end{equation} 
where $E_{n}$ and $U$ are the neutron energy and the potential, respectively. According to the hypothesis~\cite{Frank2020} in the case of accelerated motion of the potential region one should expect, in addition to the conventional Doppler change in neutron velocity, some additional effect due to acceleration. The results of the numerical calculation of the evolution of the wave packet at reflection from a potential step moving with acceleration allow such interpretation. However, the acceleration effect is strongly disguised here by a much stronger Doppler frequency shift, which makes such interpretation not very reliable. Therefore, we turn to the case of total reflection from the asymmetric potential~\cite{Bushuev2018, Frank2014} which has the form of a double step and possesses resonance properties. Under resonance conditions, the GDT at reflection increases by a factor of several orders of magnitude. It is natural to expect that the energy transfer caused by the accelerating movement of the potential will also increase significantly compared with the case of reflection from a one-step  potential. The calculations have fully met the expectations.
	
	\begin{figure*}[hbt] 
		\centering
		\captionsetup[subfigure]{labelformat=empty}
		%\hspace{-3\baselineskip}
		\begin{subfigure}[b]{0.75\columnwidth}
			\caption{}
			\vspace{-1.5\baselineskip}
			\includegraphics[width=\textwidth]{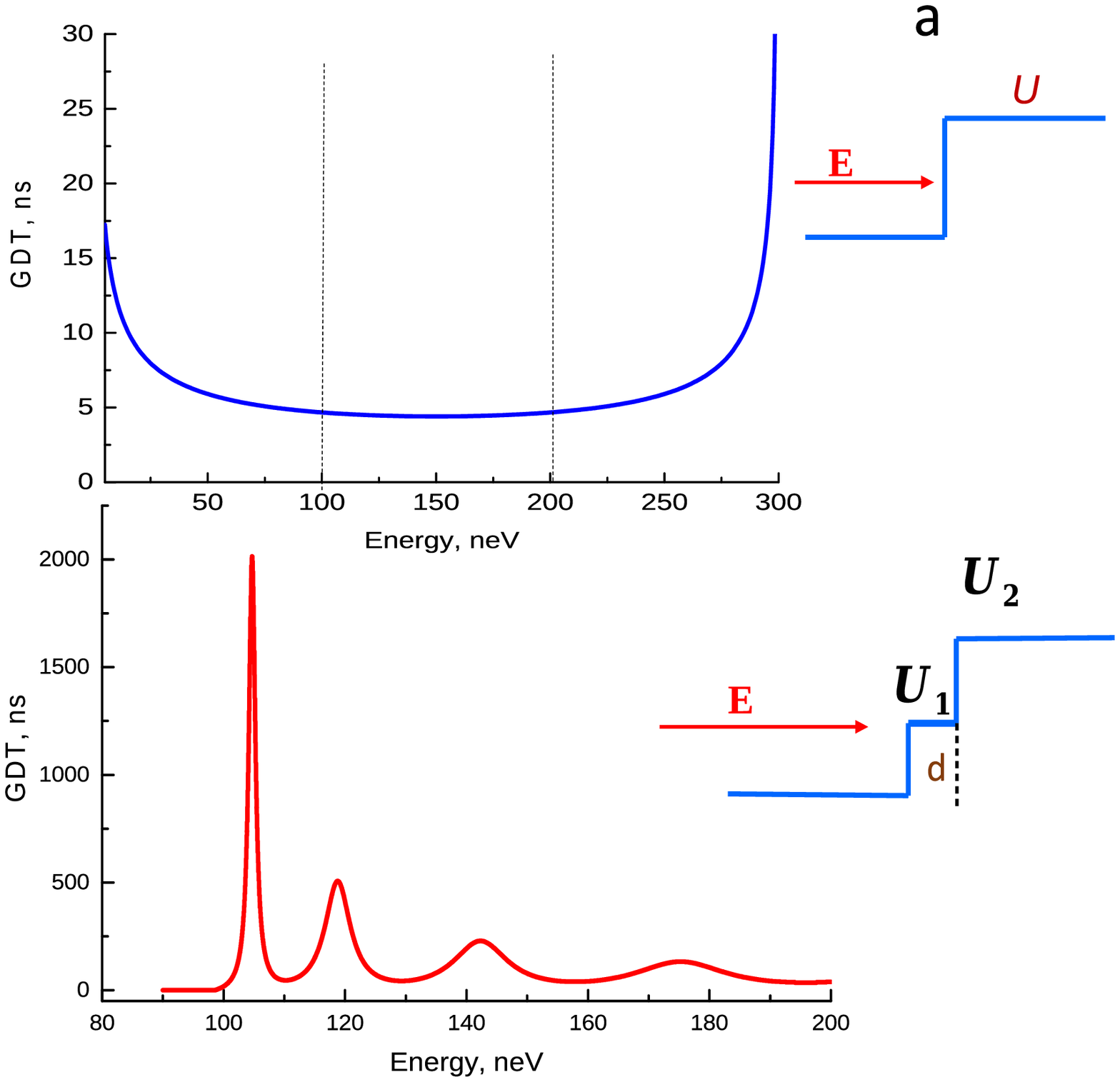}
			\label{Fig6a}
		\end{subfigure}
		% \hspace{-0.6\baselineskip}
		\begin{subfigure}[b]{0.98\columnwidth}
			\caption{}
			\vspace{-1.5\baselineskip}
			\includegraphics[width=\textwidth]{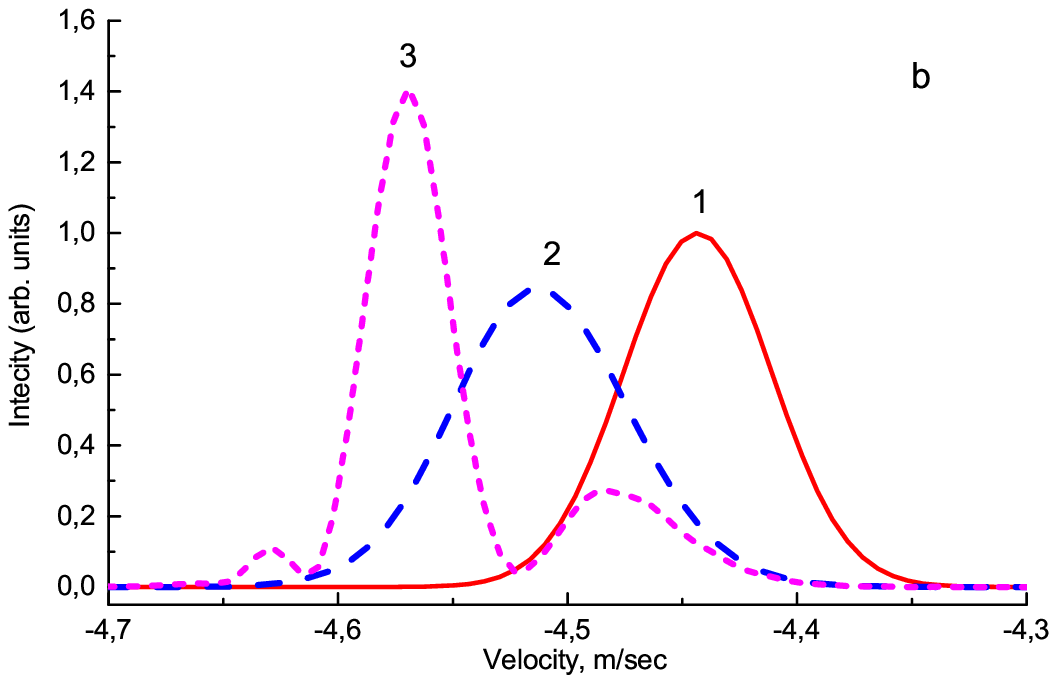}
			\label{Fig6b}
		\end{subfigure}
		\vspace{-0.7\baselineskip}
		\caption
		{The group delay time at reflection from a conventional potential barrier as well as from a barrier with an additional step (a) and the velocity spectra after reflection from a moving double-step potential (b). See the text for details.
		}
		\label{Fig6}
	\end{figure*}
	
	Figure~\ref{Fig6} illustrates the effect of acceleration at total reflection of the wave packet from a double-step resonant potential structure. Figure~\ref{Fig6a} shows the dependence of the group delay time of the neutron on the energy at its reflection from the potential step with the height $U = 300~\text{neV}$ and from the step-like potential structure characterised by the following parameters: $U_{1} = 100~\text{neV}, U_{2}= 300~\text{neV}, d = 200~\text{nm}$. The notations are indicated in the figure. Figure~\ref{Fig6b} presents three velocity spectra curves, i.e., of the wave packets in the velocity representation after reflection from a moving potential. The energy of the incident neutron slightly exceeds the potential $U_{1}$ and is equal to $E = 103.2~\text{neV}$ corresponding to the resonant delay of the reflected wave. The negative sign of the speed of the packet reflects the fact that after reflection the direction of its movement is opposite to the initial one. Curve 1 is the velocity spectrum for the packet reflected from a potential step at rest, curve 2 is the velocity spectrum of the packet reflected from a double-step potential moving with a low constant velocity towards the packet. An increase in speed as compared to case 1 is due to the Doppler effect.
	
	Curve 3 illustrates the evolution of the wave packet following its reflection from a resonant potential structure moving with the acceleration $|a| = 5\times 10^{4}~\text{m}/\text{s}^{2}$  in the opposite to neutron velocity direction. The calculation parameters are chosen such as to ensure that as soon as the wave packet maximum reaches the potential, the velocity of the potential becomes equal to the one in the absence of acceleration, i.e., the velocity presented by curve 2. 
	
	It can be seen that packet 3 is significantly deformed due to the fact that its energy width exceeds the resonance one. Therefore, different parts of the spectrum receive different increments in velocity and in energy. A significant difference in spectra 2 and 3 points to an undoubted presence of the acceleration effect at reflection from a potential structure moving with acceleration.
	
	\subsection{Resonant tunneling of neutrons through a double potential barrier - the Fabry-Perot interferometer}
	\label{subsec:Resonant tunneling of neutrons through a double potential barrier - the Fabry-Perot interferometer}
	
	One of the most interesting objects for studying the acceleration effect in quantum mechanics is the resonant potential structure, an analogue of the optical Fabry-Perot interferometer. In the practice of the neutron experiment, the simplest interferometer of the kind, also called a neutron interference filter (NIF), is a combination of three films characterized by different values of the effective potential (7). The potential structure of such a filter consists of two barriers with a well between them. With a sufficiently large width of the well $d$, formation is possible of quasi-bound states whose position is approximately determined by the relation                              
	
	\begin{equation}\label{eq21}
	k d \approx \pi, \;\;\;n = 1, 2, 3,\dots,
	\end{equation} 
where $k$ is the wavenumber corresponding to the normal component of the wave vector in the middle part of the structure - the potential well. 
	
	More accurate calculations are based on equations of continuity of wave functions at all the boundaries of such structure. An analytical solution for the reflection and transmission amplitudes of such an object is in~\cite{Xiao2012}, the problem of particle tunneling through a double-humped barrier is discussed in works~\cite{Olkhovsky2005, Balashov2004, Maaza2012}. The issue of the time of interaction with resonance structures is discussed in paper~\cite{Bushuev2018}. The work~\cite{Bjorksten1991} is devoted to the theoretical study of particle tunneling through an oscillating system of two barriers whose height oscillates with time.
	
	On the observation of resonant tunneling UCN through NIF was reported in papers~\cite{Steinhauser1980, Novopoltsev1988, Frank1999}. The physical effects that arise as ultracold neutrons pass through a NIF oscillating in space are the subject of the recent work~\cite{Zakharov2020} in which main attention is paid to non-stationary effects arising at fast modulation of the wave that passes through an oscillating potential structure. Some of the reported effects have not found any simple explanation in the work.
	
	The transmission spectrum of the interference filter is characterized by one or several narrow lines and the time of passage of a neutron through such an object exceeds the time of passage of a particle of the same path in free space by more than an order of magnitude. We would like to emphasize that we are speaking about an essentially quantum phenomenon, namely, resonant tunneling of a neutron through potential barriers. Therefore, study of the quantum problem of wave packet passage through an interference filter moving with acceleration is of considerable interest.
	
	As in the problem of wave packet reflection from a resonant structure considered above we also encounter the difficulties connected with narrowness of the resonance tunneling line as well as with an inevitable change in the NIF velocity in the course of its interaction with the wave packet. In our calculations the energy width of the wave packet was chosen to be much smaller than the width of the filter transmission line. In this case, the energy corresponding to the maximum of the spectrum of the wave packet incident on the filter coincided with the center of the resonance transmission line of the NIF at rest. 
	
	The values of the initial velocity and of the acceleration of the filter were taken on the basis of the following considerations. At the moment of time $t = 0$ the maximum of the wave packet was at a distance of $150~\mu \text{m}$ from the filter. At the same time the neutron velocity was directed in the positive direction of the $x$ axis, i.e., towards the filter while the filter itself  was moving in one of the two directions with acceleration directed opposite to its speed, i.e., with deceleration. The speed of the filter at that initial moment was large enough not to let the incident wave packet pass. The parameters of motion of the filter were such that its velocity crossed zero at the moment the packet maximum reached the filter. Note that although the velocity vector changed its sign at that, the sign of the acceleration vector remained unchanged.
	
	Thus, in the course of interaction between the wave packet and the filter the packet velocity the entire transmission region of the filter in the rest coordinate system of the filter. The evolution of the packet is traced up to the moment when it completes interaction with the potential structure with its two components corresponding to the transmitted and the reflected states having moved away from the filter at a distance of  $3\delta x$. 
	
\begin{figure}[htb]
		\centering
		\vspace{-0.3\baselineskip}
		\includegraphics[width=\columnwidth]{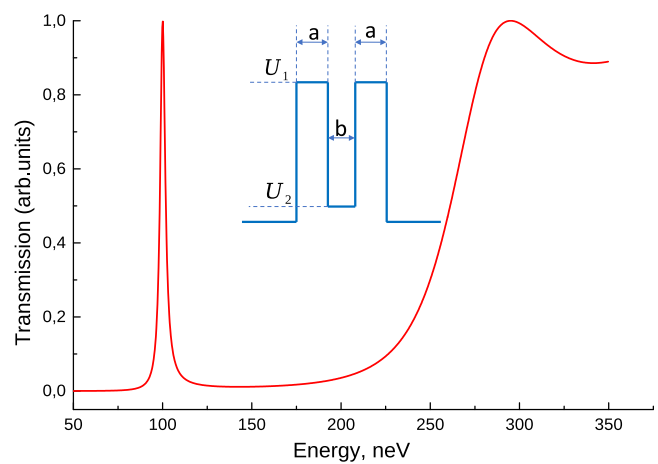}
		\caption
		{
			The potential structure of the filter and its transmission line.
		}
		\label{Fig7}
	\end{figure}

	The shape of the potential structure of the NIF and the resonant nature of its transmission are illustrated in figure~\ref{Fig7}.
	
			\begin{figure*}[htb] 
		%\centering
		%\includegraphics[width=\columnwidth]{Fig3.eps}
		%\includegraphics[width=2\columnwidth]{Fig3.eps}
		\captionsetup[subfigure]{labelformat=empty}
		%\hspace{-3\baselineskip}
		\begin{subfigure}[b]{\columnwidth}
			\caption{}
			\vspace{-1.5\baselineskip}
			\includegraphics[width=0.96\textwidth]{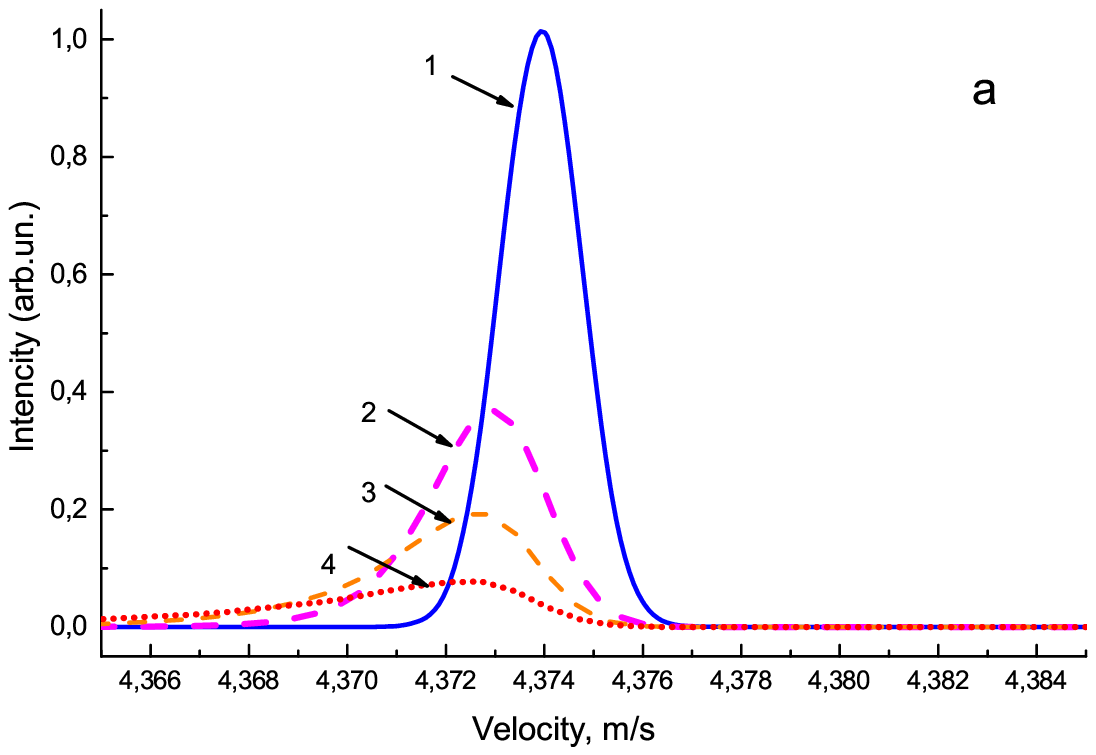}
			\label{Fig8a}
		\end{subfigure}
		% \hspace{-0.6\baselineskip}
		\begin{subfigure}[b]{\columnwidth}
			\caption{}
			\vspace{-1.5\baselineskip}
			\includegraphics[width=0.95\textwidth]{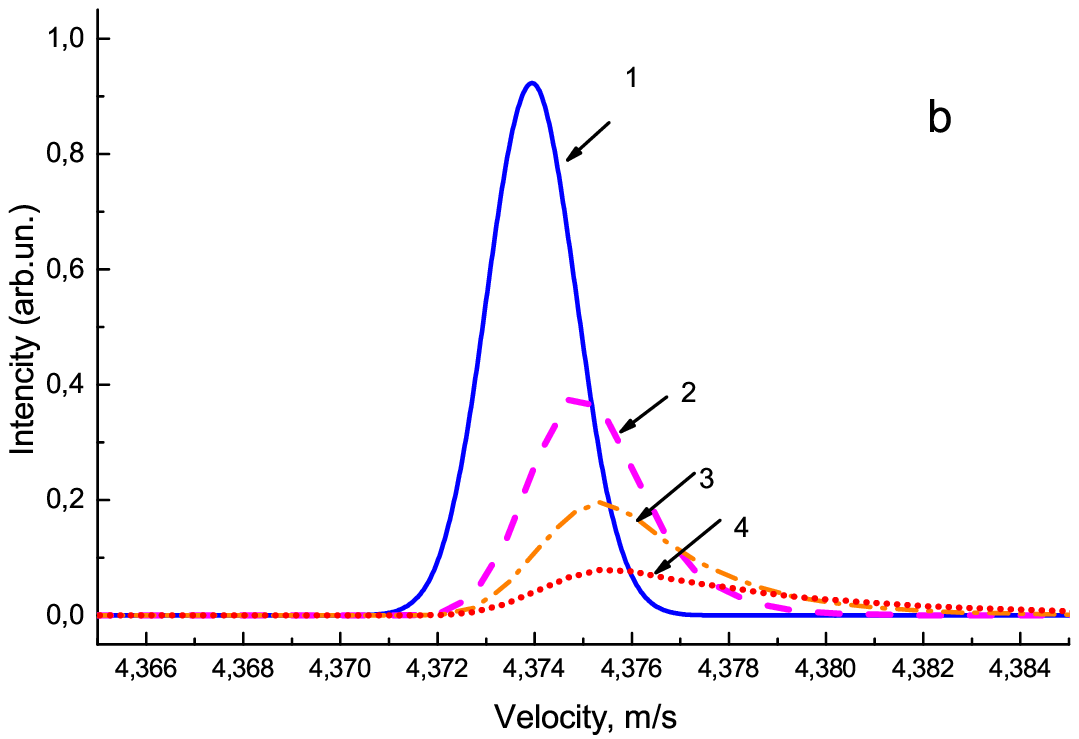}
			\label{Fig8b}
		\end{subfigure}
		\vspace{-0.7\baselineskip}
		\caption
		{The velocity spectrum of the neutron that passed through a filter moving with acceleration in the opposite to neutron velocity dirrection (a) and in the neutron velocity direction (b).}
		\label{Fig8}
		%\end{figure}
	\end{figure*}
	
	%\begin{figure}[htb]
	\begin{figure*}[htb] 
		%\centering
		%\includegraphics[width=\columnwidth]{Fig3.eps}
		%\includegraphics[width=2\columnwidth]{Fig3.eps}
		\captionsetup[subfigure]{labelformat=empty}
		%\hspace{-3\baselineskip}
		\begin{subfigure}[b]{\columnwidth}
			\caption{}
			\vspace{-1.5\baselineskip}
			\includegraphics[width=0.94\textwidth]{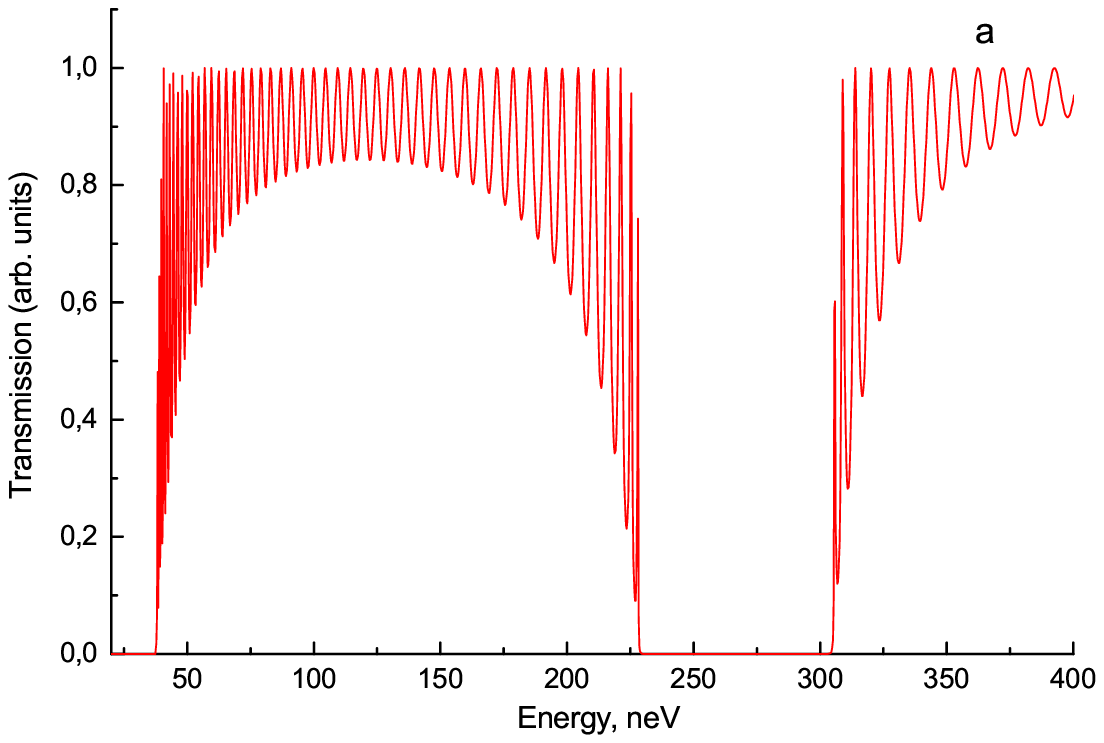}
			\label{Fig9a}
		\end{subfigure}
		% \hspace{-0.6\baselineskip}
		\begin{subfigure}[b]{\columnwidth}
			\caption{}
			\vspace{-1.5\baselineskip}
			\includegraphics[width=0.98\textwidth]{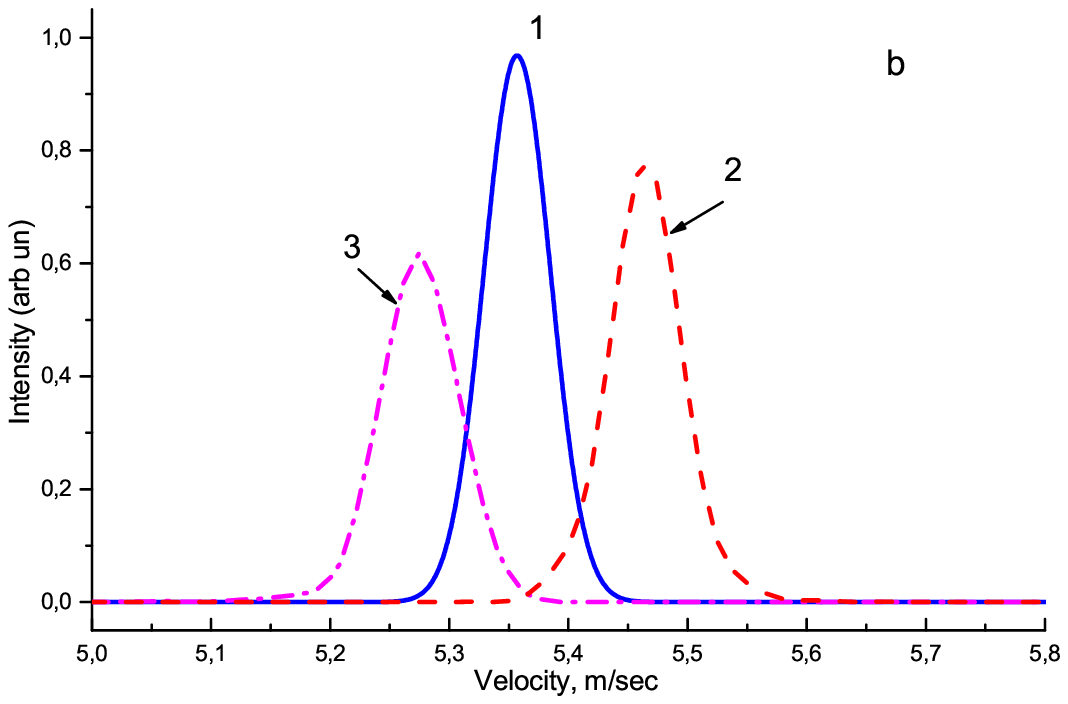}
			\label{Fig9b}
		\end{subfigure}
		\vspace{-0.7\baselineskip}
		\caption
		{The dependence of the transmission through a periodic potential structure on the neutron energy (a) and the evolution of the velocity spectra as neutrons pass through the  structure moving with acceleration (b).}
		\label{Fig9}
		%\end{figure}
	\end{figure*}

	The parameters of the potential structure are taken to be: $U_{1} = 200~\text{neV}, U_{2} = 2.15~\text{neV}, a = 30~\text{nm}, b = 23~\text{nm}$. The maximum of the transmission line corresponds to the energy of $E = 100~\text{neV}$ and the dispersion of the wave packet in the coordinate space is $\delta x = 40 ~\mu \text{m}$.
	
		Figure~\ref{Fig8} shows the velocity spectra (wave packet shapes) of states that have passed through a filter. The numbers on the curves correspond to the following parameters: (1) is the spectrum of the initial state, (2) is the spectrum at absolute value of acceleration $|a| = 5\times 10^{3}~\text{m}/\text{s}^{2}$, (3) is the spectrum at $|a| = 1\times 10^{4}~\text{m}/\text{s}^{2}$, (4) is the spectrum at $|a| = 2\times 10^{4}~\text{m}/\text{s}^{2}$. 
	
	From comparison of the curves presented in figure~\ref{Fig8} it follows that the spectrum of the neutrons transmitted through the filter does shift, with the shift being dependent on the sign and magnitude of acceleration. In this case, there occurs the smearing of the spectrum that also depends on the magnitude of acceleration. This is possibly due to a noticeable change in the filter velocity as the packet interacts with a resonant potential structure. Thus, the acceleration effect obviously exists in the case of quantum tunnelling as well.	 	
	
	\subsection{Neutron tunnelling through a regular system formed by potential barriers}
	\label{Neutron tunnelling through a regular system formed by potential barriers}
	
	Having demonstrated the existence of an acceleration effect for tunnelling through a simplest resonant system in a numerical experiment, it was natural to turn to the case of a more complex potential structure consisting of a sufficiently large number of potential barriers. The fact that such a model is connected with a large family of multilayer interference mirrors and photonic crystals that are widely used in conventional optics, x-ray or neutron optics is quite evident.
	
	As a model we take a system of 51 potential barriers 25 nm wide 5 nm apart.  The barriers are 250 neV high. Figure~\ref{Fig9a} shows the dependence of transmission through such a structure on the neutron energy. It can be seen that in the energy range from 228 to 305 neV, which corresponds to Bragg reflection, the structure does not transmit neutrons. It is important that for neutron energies below the potential barrier there exists rather a wide range of energies over which the transmission is significant. We emphasise that the nature of such transmission is close to the nature of resonant tunnelling through a simple interference filter with a single potential well. To some extent, a periodic structure with many barriers and wells can be considered as a combination of many elementary resonators. A continuous barrier with an equal thickness is absolutely non-transparent. The GDT value in the region of the highest transparency of the structure oscillates slightly remaining in the range 350 - 450 ns.
	
	Figure~\ref{Fig9b} shows the transmission spectrum of the structure in the absence of acceleration and in the case of the structure moving with acceleration in two opposite directions. The numbers on the curves correspond to the following: 1 - the spectrum of the initial state, 2 - the acceleration in the direction of neutron velocity $a = 5\times 10^{5}~\text{m}/\text{s}^{2}$, 3 - the acceleration in the direction opposite to neutron velocity $a = - 5\times 10^{5}~\text{m}/\text{s}^{2}$.
		
	\section{Conclusion}
	
	The results of the numerical study of the problem of interaction of ultracold neutrons with potential structures moving with constant acceleration are presented. The solution of the time-dependent Schrodinger equation was found by the method of evolution operator splitting. The problem of neutron passage above an accelerating potential barrier was considered for the case of the particle energy significantly higher than the barrier, in which the role of interference effects is negligible, and for the case of little difference in the energies. The first allows a simple analysis on the basis of a classical approach, which made it possible to confirm with a high degree of reliability the correctness of the numerical method used.
	
	The case of total reflection of a neutron from a double potential step moving with acceleration was investigated. This model was applied because at reflection from such a structure the GDT significantly exceeds the reflection time from a simple one-step potential. Among the problems considered there were also two problems connected with the spectra of neutrons arising after tunneling through a potential structure moving with acceleration. In one case, such a structure consisted of two barriers with a well between them, which is characteristic of the simplest resonator - the Fabry Perot interferometer. In the other case, the potential structure was a succession of alternating barriers and wells, which is typical of multilayer mirrors - monochromators.
	
	In the calculation, the shape of the wave packet after interaction with an accelerating object was compared with that of the initial packet. In all the cases, the interaction resulted in a shift in the speed distribution accompanied, in some cases, by the broadening of the spectrum. The change of the most probable velocity, the packet maximum in the velocity representation, was more or less proportional to acceleration and changed its sign as the direction of acceleration reversed. The change in the velocity corresponded, in the order of magnitude, to the relation $\Delta \langle v \rangle = a \tau$   where $\langle v \rangle$  is the average velocity and $\tau$ is the group delay time~\eqref{eq6}. No precise fulfillment of the condition was probably to be expected as the parameters of the problem changed noticeably in the course of the wave packet passage through the region of the potential due to high acceleration.
	
	The results obtained are consistent with the qualitative prediction~\cite{Frank2020} concerning changes in the frequency of the wave and in the energy of the particle as a result of its interaction with an accelerating object not only in the case of optical but also in the case of essentially quantum phenomena. Consequently, the hypothesis that the acceleration effect supplementing the usual Doppler effect is of a very general nature is confirmed.
	
	Let us make two more comments here. In the above discussed problems all the calculations were performed for a particle with the parameters characteristic of UCNs. This was due to the fact that it is UCNs that are most suitable for experimental verification of the validity of the obtained results that lie far beyond the scope of neutron optics and are of a much more general character.
	
	The second comment concerns the fact that above we assumed that the concept of the effective potential of matter is valid (1), including the case of high acceleration of the sample. At the same time, in the case of a medium moving with high acceleration, the validity of such an assumption is not obvious at all~\cite{Frank2006, Frank2008, Frank2014JETP}. Moreover, if the acceleration effect is really universal, it should also fully apply to the case of neutron scattering on an atomic nucleus moving with acceleration. This cannot but entail some correction of our ideas about the dispersion law for neutron waves propagating in matter moving with acceleration.
%---------------------------------------------------
	\begin{acknowledgement}	
		We are incredibly grateful to Prof. V.A. Bushuev for fruitful discussions.
	\end{acknowledgement}
%
% BibTeX users please use

 \bibliographystyle{spphys}
 \bibliography{Refs.bib}

\end{document}